\documentclass[aps,prb,showpacs,twocolumn,superscriptaddress,floatfix]{revtex4-1}
\usepackage{setspace}
\usepackage{array,tabularx,booktabs}
\usepackage{appendix}

\usepackage{graphicx}
\usepackage{epstopdf}
\usepackage{color}
\usepackage[colorlinks,bookmarks=false,citecolor=blue,linkcolor=red,urlcolor=blue]{hyperref}
\usepackage{tikz}
\usetikzlibrary{calc}
\usepackage{multirow}
\usepackage{bm}

\usepackage{amsmath,amssymb}
\usetikzlibrary{shapes}

\DeclareRobustCommand{\uuu}{
\left|\uparrow\uparrow\uparrow\right>
}

\DeclareRobustCommand{\duu}{
\left|\downarrow\uparrow\uparrow\right>
}

\DeclareRobustCommand{\udu}{
\left|\uparrow\downarrow\uparrow\right>
}

\DeclareRobustCommand{\uud}{
\left|\uparrow\uparrow\downarrow\right>
}

\DeclareRobustCommand{\udd}{
\left|\uparrow\downarrow\downarrow\right>
}

\DeclareRobustCommand{\dud}{
\left|\downarrow\uparrow\downarrow\right>
}

\DeclareRobustCommand{\ddu}{
\left|\downarrow\downarrow\uparrow\right>
}

\DeclareRobustCommand{\ddd}{
\left|\downarrow\downarrow\downarrow\right>
}

\DeclareRobustCommand{\allout}{
\begin{tikzpicture}[baseline={([yshift=-\the\fontdimen22\textfont2]current bounding box.center)}]]
       \filldraw [color=gray,fill=white,thick ,scale=0.7] (0., 0.)--(0.4, 0.)--(0.2,0.346)--cycle;
 \draw [->,scale=0.7] (0.2, 0.19) -- (0.2, 0.49);
 \draw [->,scale=0.7] (0.27, 0.075) -- (0.53,-0.075);
  \draw [->,scale=0.7] (0.13,0.075) -- (-0.13, -0.075);
   \end{tikzpicture}
     }

\DeclareRobustCommand{\allin}{
\begin{tikzpicture}[baseline={([yshift=-\the\fontdimen22\textfont2]current bounding box.center)}]]
       \filldraw [color=gray,fill=white,thick, scale=0.7] (0., 0.)--(0.4, 0.)--(0.2,0.346)--cycle;
 \draw [<-, scale=0.7] (0.2, 0.19) -- (0.2, 0.49);
 \draw [<-,scale=0.7] (0.27, 0.075) -- (0.53,-0.075);
  \draw [<-, scale=0.7] (0.13,0.075) -- (-0.13, -0.075);
   \end{tikzpicture}
     }

\begin{document}

\title{Multipolar edge states in the breathing kagom\'e model}

\author{Judit Romh\'anyi}
\affiliation{Okinawa Institute of Science and Technology Graduate University, Onna-son, Okinawa 904-0395, Japan}
\date{\today}

\begin{abstract}

Excitations of ordered insulating magnets gain renewed interest due to their potential topological properties and the natural realization of magnetic analogues of the celebrated topological models. In this paper we go beyond these parallels 
and explore what else is there in the unconventional excitations of quantum magnets.  
We study the topologically nontrivial multiplet excitations of the antiferromagnetic spin-$1/2$ kagom\'e system with strong breathing anisotropy and Dzyaloshinskii-Moriya interaction. We show that in the chiral magnetic ground state the excitations can be characterized by a spin-$1/2$ doublet and a spin-$3/2$ quartet. 
With the use of magnetic field we can tune the quartet through a band touching topological  phase transition, when a novel spin-3/2 Dirac cone is formed by the touching of four bands.
In the topologically nontrivial regime the spin-$3/2$ bands have large Chern numbers $-3$, $-1$, $1$, $3$.  In an open system the emerging chiral edge states naturally inherit the multipolar characters and we find novel quadrupolar edge modes.

\end{abstract}


\maketitle

\section{Topological magnetic excitations}
Since the discovery of quantum Hall effect the concept of topology has rapidly spread from electronic systems to a wider scope including bosonic excitations.
In addition to the Hall effect of photonic crystals~\cite{Raghu2008,Petrescu2012,Rechtsman2013,Hafezi2013,Khanikaev2013,Lu2014}, phonon modes~\cite{Zhang2010,Zhang2011,Qin2012} and skyrmion textures~\cite{Hoogdalem2013,Jiang2017,Litzius2017}, the
 topological aspect of magnetic excitations in insulating quantum magnets continues to attain growing interest.
The  topological Haldane model realized naturally in the magnon spectrum of iron-based honeycomb insulators~\cite{Kim2017}
 and the Weyl magnons emerging in breathing pyrochlore lattice~\cite{Li2016} illustrate well the advances of magnetic analogues of topological electron bands.  
A great advantage, beside the natural realization of these modes in the magnon spectrum of ordered magnets, is their  easy access and control by magnetic field.  
 
Here we propose, that the unconventional excitations of simple ordered quantum magnets can not only hold similar properties to the topological electron bands, but can offer a new playground for novel multipolar topological bands arising naturally as a consequence of quantum mechanical entanglement between magnetic degrees of freedom.

Theoretical prediction of the Hall effect of topologically nontrivial triplet excitations~\cite{Romhanyi,Malki2017} and the experimental observation of topologically protected chiral edge modes in the orthogonal dimer system, SrCu$_2$(BO$_3$)$_2$~\cite{McClarty} opened a route to generalization of topological magnons. Expanding this idea, we propose that two-dimensional quantum magnets with breathing anisotropy can exhibit topologically nontrivial multiplet excitations similar to the triplets observed in SrCu$_2$(BO$_3$)$_2$. Considering a  cluster of spins, the local Hilbert space is enlarged, naturally encompassing higher multipole degrees of freedom. A plaquette of three $S=1/2$ quantum spins, for example, gives room for an effective spin-$3/2$ quartet, the next multiplet after the spin-$1$ triplet of the dimerized system.

We study the breathing  kagom\'e  lattice with $S=1/2$ spins in the framework of weakly interacting trimers. Keeping these triangular building blocks entangled, the excitation spectrum naturally includes propagating multiplets.
We show that the $S=3/2$ quartet excitations have nontrivial topology, and due to their multiplet nature, large Chern numbers and corresponding multipolar edge states.

\begin{figure}[h]
\begin{center}
\includegraphics[width=0.8\columnwidth]{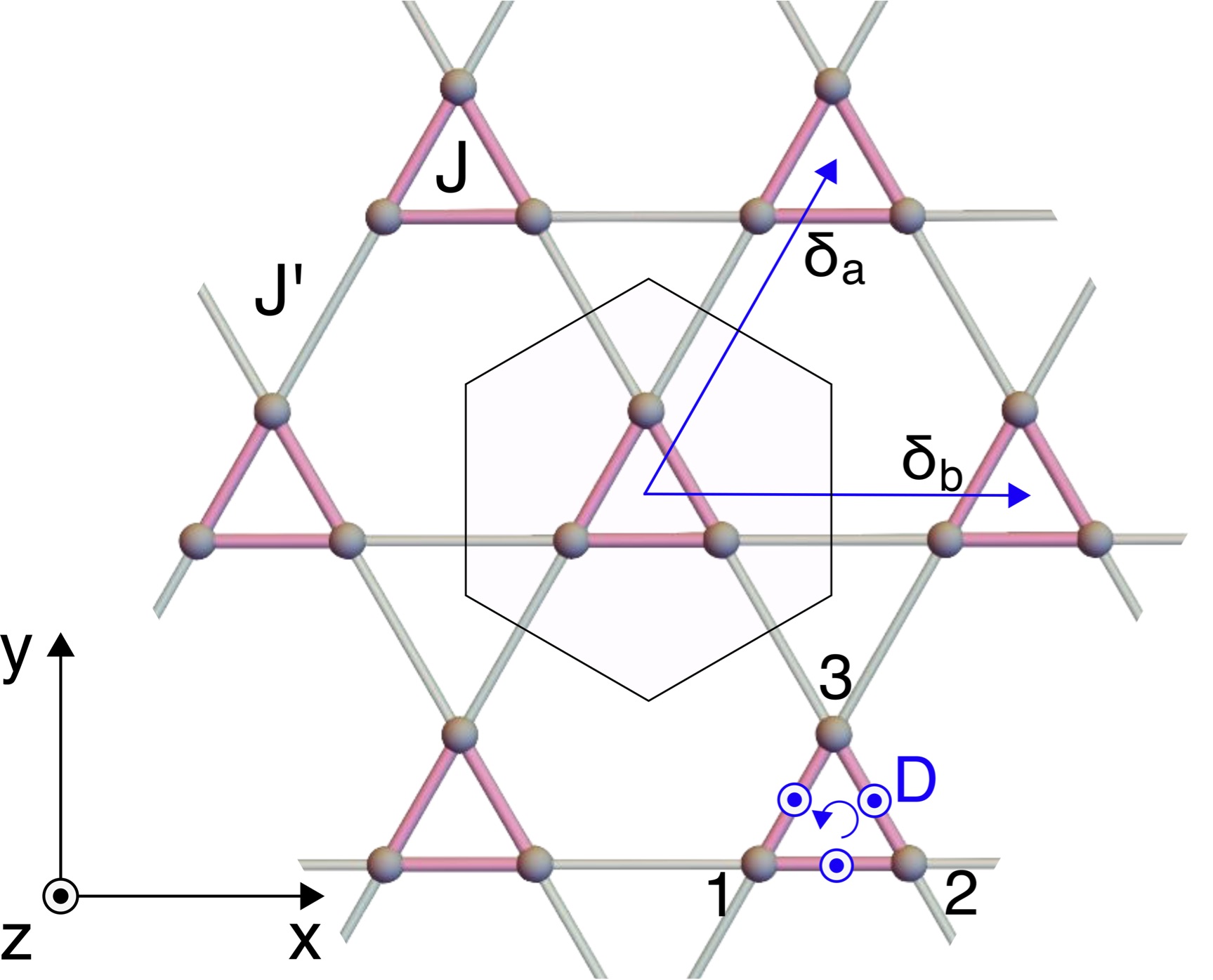}
\caption{
Breathing kagom\'e lattice with Heisenberg and DM couplings. The intra-trimer DM vector $\mathbf{D}=\!(0,0,D)\!$ is pointing in or out of the lattice plane on each bond as we go around the strong up-triangles indicated by blue arrow. The coinciding structural and magnetic unit cell is shown by a hexagon, together with the lattice translation vectors $\delta_a$ and $\delta_b$.
}
\label{fig:lattice}
\end{center}
\end{figure}

\section{Results}
\subsection{Model Hamiltonian}

We consider the trimerized antiferromagnetic kagom\'e model with intra-trimer Dzyaloshinskii-Moriya (DM) anisotropy and magnetic field perpendicular to the lattice plane. Fig~\ref{fig:lattice} illustrates the breathing kagom\'e lattice together with the magnetic and coinciding crystallographic unit cell. The Hamiltonian is given by Eq.~\ref{eq:Hamilton}.
\begin{eqnarray}
\mathcal{H}&=& J \sum_{i,j\in\bigtriangleup} \boldsymbol{S}_i\cdot\boldsymbol{S}_j
+J' \sum_{i,j\in\bigtriangledown} \boldsymbol{S}_i\cdot\boldsymbol{S}_j\nonumber\\
&&+D\sum_{i,j\in\bigtriangleup} (\boldsymbol{S}_i\times\boldsymbol{S}_j)_z-g_z h^z\sum_i S^z_i\;.
\label{eq:Hamilton}
\end{eqnarray}

\subsection{Ground state and the gapless excitation}

As a first step we construct an appropriate basis for the weakly interacting trimers. A natural starting point is to diagonalize the eight-dimensional local Hilbert space of a single triangle. The SU(2) symmetry splits this basis into two four-dimensional subspaces as $\mathcal{D}^{(\frac{1}{2})}\otimes \mathcal{D}^{(\frac{1}{2})}\otimes \mathcal{D}^{(\frac{1}{2})}=2\mathcal{D}^{(\frac{1}{2})}\oplus \mathcal{D}^{(\frac{3}{2})}$. Therefore, when spin rotational symmetry is preserved, and in case of antiferromagnetic exchange, the spectrum of a single triangle consists of two low-lying spin-$1/2$ doublets and a high-energy spin-$3/2$ quartet. The intra-trimer DM interaction is, in fact, the vector chirality of the three spins forming a trimer. Taking the $z$-component of DM into account, the spin rotation symmetry is broken down to U(1), and the low-lying doublets split as $D$ selects a specific chirality. The non-chiral $S=\frac 3 2$ states remain unaffected by the DM coupling. A finite magnetic field further lowers the symmetry and Zeeman-splits the multiplets. 
We use the spin, its $z$-component, and the $z$-component of the vector chirality to distinguish the states, and introduce the notation $\left|S,m,\chi^z_v\right>$ to label them. 
Details of the multiplet basis are given in the Supplementing Materials (SM)~\ref{sec:basis}.

In the following, we fix the sign of $D$ to be negative. The spectrum of the isolated triangles can be divided into three parts: 
A large gap ($\sim \frac{3J}{2}$) separates the non-chiral quartet and the chiral doublet manifolds, furthermore the DM interaction shifts the spin-half states with positive and negative chirality into the opposite direction:
$D<0$ lowers the energy of states with positive chirality, $\left|S,m,\chi^z_v\right>=\left|\frac 1 2,\pm\frac 1 2, +\right>$. These are degenerate when the magnetic field is zero and form the lowest lying subspace. Taking the inter-trimer coupling $J'$ into account, the $\pm\frac 1 2$ states become mixed to minimize the energy of the inter-trimer bonds.
We write the new basis for the lowest-lying doublet as 
\begin{eqnarray}
&&\textstyle{
\left|\allout\right>=\cos\vartheta \left|\frac 1 2,\frac 1 2,+\right> +e^{i \varphi} \sin\vartheta \left|\frac 1 2 ,\!-\!\frac 1 2,+\right>}\nonumber\\
&&\textstyle{\left|\allin\right>=\sin\vartheta \left|\frac 1 2,\frac1 2,+ \right> -e^{i \varphi} \cos\vartheta \left|\frac 1 2,\!-\!\frac 1 2,+\right>}\;,
\end{eqnarray}

where the variational parameter $\varphi$ can take arbitrary value reflecting the remaining U(1) symmetry.   For convenience, we choose $\varphi=-\frac{5 \pi}{6}$, so that in the ground state the spins are pointing all out on the up-triangles (and all in for the orthogonal state).  $\vartheta=\frac 1 2\arccos\frac{g_z h^z}{J'}$ is obtained by minimizing the ground state energy, 
\begin{equation}
\textstyle{E_0=\!-\frac{3J}{4}\!\!+\!\frac{\sqrt{3}D}{2}\!-\!\frac{J'(1\!-\!3\cos4\vartheta)}{24}\!-\!\frac{ g_z h^z\cos2\vartheta}{2}.}\label{eq:mf_en}
\end{equation}
The parameter $\vartheta$ determines how much the spins are tilting out of the plane towards the magnetic field. 
When the field is zero, $\vartheta=\frac{\pi}{4}$, and the $\pm\frac 1 2$ states mix equally to form a planar $120^\circ$-type of order. A  finite magnetic field cants the spins uniformly out of the plane and a chiral magnetic `umbrella'  state is realized as the ground state.
Let us note that the ground state is not the classical planar or canted $120^\circ$ state. The trimers are entangled which is manifested in the shorter spin length; 
\begin{equation}
\textstyle{\left<\allout\right|\boldsymbol{S}_n\left|\allout\right>=R_z(n\frac{2\pi}{3})\left(0,\frac 1 3\sin2\vartheta,\frac 1 6\cos2\vartheta\right)}\;,
\end{equation}
with $n=1,2,3$ denoting the sites of the triangle and the operator $R_z(\alpha)$ corresponding to a real $\alpha$-rotation about the $z$-axis. 
The spin expectation value for the orthogonal all-in state $\left|\allin\right>$ differs only in sign.
We note that $J'$ couples the lowest all-in all-out subspace to the states $\left|3/2,\pm3/2,0\right>$, too. However, the mixing with the $3/2$ multiplet is second order in $J'/J$ and negligible compared to the mixing within the spin-half subspace. This is not surprising considering the energy gap separating the spin-$3/2$ and $1/2$ multiplets. 

\begin{figure*}[ht!]
\includegraphics[width=6in]{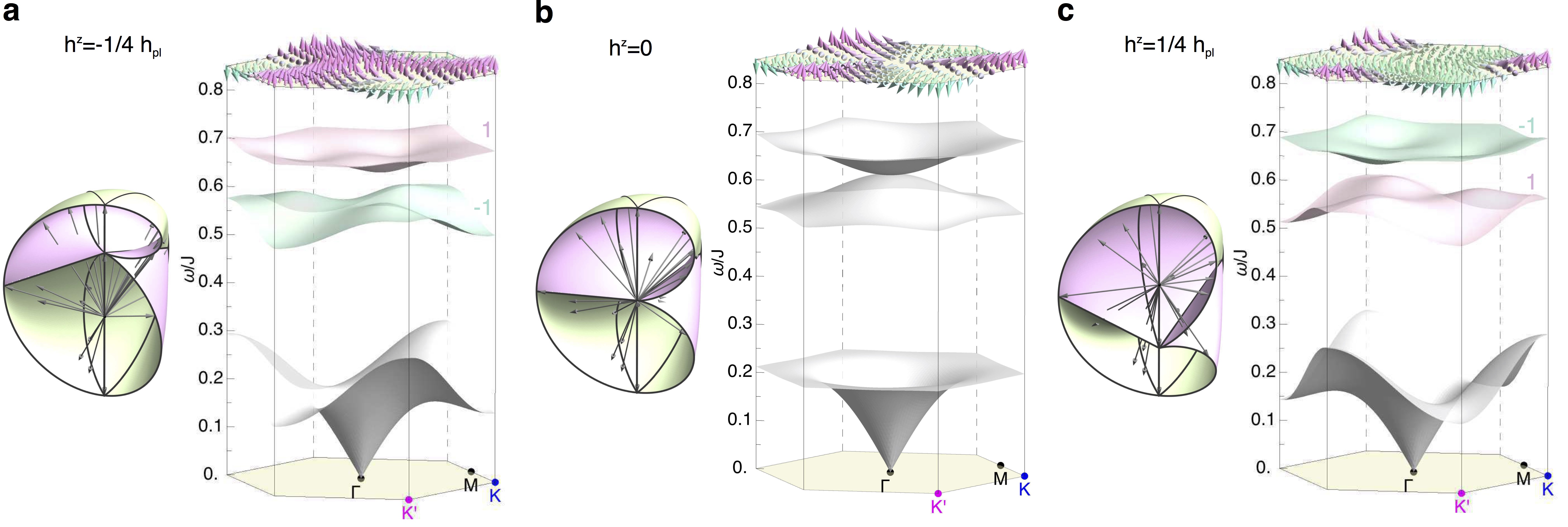}
\caption{
Topological bands of the doublet subspace. The lowest, topologically trivial mode corresponding to the propagation of  $\left|\allin\right>$  and given by $\omega_{\bf k}$ of Eq.~\ref{eq:golsdstone} is displayed in gray color.  (a) At  $h_{\sf pl}<h^z<0$ fields the doublet excitations are non-degenerate and the Chern number is well defined taking the values $-1,1$. The pseudo magnetic field ${\boldsymbol d}_{\bf k}$ forms a skyrmion in the hexagonal BZ as illustrated above the dispersions. Drawing these vectors in a common origin, they map out a surface plotted next to the bands. For better visualization we cut these surfaces in half, so that one can see inside. When the origin is enclosed in the surface, the topology of the doublet is non-trivial. (b) At $h^z=0$ the gap closes at the $\Gamma$ point, indicating that ${\bf d}_{\Gamma}={\bf 0}$, and the origin is on the surface between the chambers. The flux of the ${\boldsymbol d}_{\bf k}$ is zero. (c) For $h^z>0$, the gap opens again, the origin moves to the upper chamber of the surface and the topology of the doublets is reversed with the reversed flux of the pseudo-magnetic field ${\boldsymbol d}_{\bf k}$ illustrated above the dispersion. More information on the skyrmions of ${\boldsymbol d}_{\bf k}$ can be found in the SM. ($J=1$, $J'=0.3$, $D_z=-0.3$, and $g_z=2$)}
\label{fig:topo_1o2}
\end{figure*}

At a trimer-product level, the ground state is the realization of $\left|\allout\right>$ on each strong triangle of the breathing lattice. We project the full Hamiltonian~\ref{eq:Hamilton} onto the lowest-lying subspace, and treat $\left|\allout\right>$ and $\left|\allin\right>$ as the components of a pseudo-spin. Using conventional linear spin wave theory, we calculate the lowest excitation which can be imagined as `flipping' a pseudo-spin, i.e. removing a state $\left|\allout\right>_j$ and creating $\left|\allin\right>_j$. Such a local excitation can propagate over the lattice via $J'$ to lower the kinetic energy. 
We introduce a boson for this excitation: $\left|\allin\right>_j=a^\dagger_j\left|0\right>$, where the vacuum state, $\left|0\right>$, corresponds to the condensation of $\left|\allout\right>$.
The spin wave Hamiltonian has the following form
\begin{eqnarray}
\mathcal{H}^{\rm sw}=\!\!\left(\!\!
\begin{array}{c}
a^{\dagger}_{\bf k}\\
a^{\phantom{\dagger}}_{-\bf k}
\end{array}\!
\right)\!\!
\left(\!\!
\begin{array}{cc}
M^{\phantom{\dagger}}_{\bf k}& N^{\phantom{\dagger}}_{\bf k}\\
N^{^{\phantom{\dagger}}*}_{-{\bf k}} & M^{^{\phantom{\dagger}}*}_{-{\bf k}}
\end{array}\!\!
\right)
\!\left(\!\!
\begin{array}{c}
a^{\phantom{\dagger}}_{\bf k}\\
a^{\dagger}_{-\bf k}
\end{array}\!
\right)\!\!\;,
\label{eq:lsw}
\end{eqnarray}
with $M_{\bf k}=\frac{J'}{6}\left[4\!-\left(1+\frac{3(g_zh^z)^2}{J'^2}\right)\!\gamma_1\right]+\frac{2\sqrt{3}}{3}g_zh^z\gamma_2$ and $N_{\bf k}=\frac{J'}{2}\left[1-\frac{(g_zh^z)^2}{J'^2}\right]\gamma_1$. %
Eq.~\ref{eq:lsw} is diagonalized using Bogoliubov transformation. The dispersion of the resulting mode is given by
\begin{equation}
\omega_{\bf k}=\textstyle{\frac{J'}{3}\sqrt{2(1-\gamma_1)\left[(2+\gamma_1)-\frac{3(g_zh^z)^2}{J'^2}\right]}+\frac{2\sqrt{3}}{3}g_zh^z\gamma_2}\;.
\label{eq:golsdstone}
\end{equation}
and is shown in Fig.~\ref{fig:topo_1o2}. The geometrical factors are 
\begin{eqnarray}
\gamma_1=\frac{1}{3}\sum^3_{n=1}\cos{\boldsymbol k}\cdot{
\boldsymbol \delta}_n\quad\text{and}\quad
\gamma_2=\frac{1}{3}\sum^3_{n=1}\sin{\boldsymbol k}\cdot{
\boldsymbol \delta}_n\;,
\label{eq:geom_fact}
\end{eqnarray}
 where $\boldsymbol{\delta}_n$ can take the values $\boldsymbol{\delta}_1=\boldsymbol{\delta}_b-\boldsymbol{\delta}_a=(\!-\! 1/2,\sqrt{3}/2)$, $\boldsymbol{\delta}_2=\boldsymbol{\delta}_a=(1/2,\sqrt{3}/2)$ and $\boldsymbol{\delta}_3=-\boldsymbol{\delta}_b=\left(-1,0\right)$ as introduced in Fig.~\ref{fig:lattice}. Eq.~\ref{eq:lsw} is derived in the SM~\ref{sec:basis}.
 
Increasing the magnetic field to $h_{\rm pl}=\pm J'/g_z$, $\omega_{\bf k}$ softens quadratically at the $K/K'$ point of the Brillouin zone (BZ) and a symmetry breaking phase transition takes place from the chiral  state into a gapped uniform 1/3-plateau phase above $h_{\rm pl}$. The plateau phase is characterized by $\vartheta=0$, $\boldsymbol{S}_j=\left(0,0,\frac 1 6\right)$ and the condensation of $\left|\frac 1 2 ,\!-\!\frac 1 2,+\right>$ state on each up-triangle.

\subsection{Topology of multiplet excitations}

  The remaining six orthogonal states; the doublet with negative chirality, $|\frac 1 2,m,-\rangle$ and the non-chiral quartet, $|\frac 3 2,m,0\rangle$ comprise the  gapped excitations. 
The multiplet dynamics in leading order in $J'/J$ is characterized by hopping processes such as 
\begin{equation}
{}_i\langle \frac 1 2, m,-|\mathcal{H}|\frac 1 2,m', -\rangle_{j}\quad\text{and}\quad{}_i\langle \frac 3 2, m,0|\mathcal{H}|\frac 3 2,m', 0\rangle_{j}
\end{equation}
where the site indices $i$ and $j$ are either the same or belong to neighboring up-triangles, and $m$ and $m'$ can take the values $-S,\hdots,S$, for $S=\frac 1 2$ and $\frac 3 2$.

Introducing bosonic operators $\left|\frac 1 2,m,-\right>_j=b^\dagger_{m,j}\left|0\right>$, and $\left|\frac 3 2,m,0\right>_j=c^\dagger_{m,j}\left|0\right>$  for the two multiplets, we calculate the hopping matrix elements. (see SM~\ref{sec:effective_1o2} and~\ref{sec:effective_3o2}) After Fourier transformation, the effective Hamiltonians is obtained in the form of
\begin{eqnarray}
&&\mathcal{H}^{(1/2)}=\sum_{k}\sum_{mm'}b^\dagger_{m,{\bf k}} B_{mm'}({\bf k})b^{\phantom{\dagger}}_{m',{\bf k}}\quad\text{and}\nonumber\\
&&\mathcal{H}^{(3/2)}=\sum_{k}\sum_{mm'}c^\dagger_{m,{\bf k}} C_{mm'}({\bf k})c^{\phantom{\dagger}}_{m',{\bf k}}\;,
\label{eq:hopping}
\end{eqnarray}
where $B_{mm'}({\bf k})$ and $C_{mm'}({\bf k})$ are $2S+1$-dimensional matrices with $S=1/2$ and $3/2$, respectively. 
We can rewrite the hopping matrices in the convenient form
\begin{subequations}
\begin{eqnarray}
&&B({\bf k})=\Delta^{(1/2)}_{\bf k}\cdot\boldsymbol{1}_{2}+\boldsymbol{d}^{(1/2)}_{\bf k}\cdot\boldsymbol{s}\quad\text{and}\\
&&C({\bf k})=\Delta^{(3/2)}_{\bf k}\cdot\boldsymbol{1}_{4}+\boldsymbol{d}^{(3/2)}_{\bf k}\cdot\boldsymbol{S}\;,
\label{eq:M_matrix}
\end{eqnarray}
\label{eq:dirac}
\end{subequations}
where $\boldsymbol{1}_{n}$ is an $n$-dimensional unit matrix, $\Delta_{\bf k}^{(S)}$ is a shift in the energy corresponding to the gap and $\boldsymbol{d}^{(S)}_{\bf k}$ is a pseudo magnetic field that induces a Zeeman-like splitting of the $S$ multiplets.
For the $S=1/2$ subspace, the vector $\boldsymbol{s}$ simply contains the three Pauli matrices, $\frac 1 2\boldsymbol{\sigma}$, while for the quartet, $\boldsymbol{S}$ corresponds to the $x,y,z$-components of a spin-$3/2$ quantum spin.

In case of the doublet, obtaining the simple form of $\Delta^{(1/2)}_{\bf k}\cdot\boldsymbol{1}_{2}+\boldsymbol{d}^{(1/2)}_{\bf k}\cdot\boldsymbol{s}$ is obvious, as the operator space of a quantum spin-$1/2$ is four-dimensional, and $\boldsymbol{s}$ together with  $\boldsymbol{1}_{2}$ form a basis for it. 

This argument breaks down in case of the quartet, as the operator space is $16$-dimensional, including seven octupole and five quadrupole operators beside the spin components, $\boldsymbol{S}$ and the identity operator, $\boldsymbol{1}_{4}$ which contribute to the  dynamics of the quartet. 
%
\begin{figure*}[ht!]
\includegraphics[width=7in]{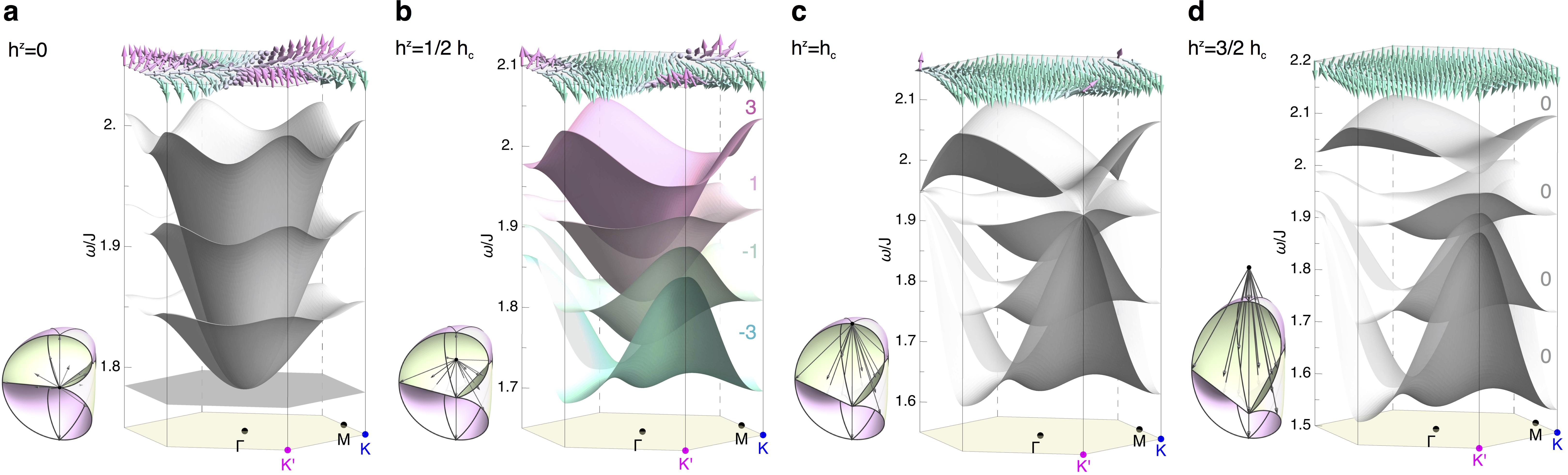}
\caption{
Topological band touching transition in the quartet subspace. Next to the dispersions we show the surface formed by the pseudo-field $\boldsymbol{\delta}_{\bf k}$. For better understanding, we cut them in halves, to reveal the inside chambers and the position of the origin. (a) At zero field there is quadratic touching at the $\Gamma$ point. For $0<h^z<h_c$ fields (b) the  bands become fully split and the Chern number is a well defined quantity with the values $2m=3,1,-1,-3$. (c) At the critical field, $h^z=h_c$, the bands undergo a band-touching topological transition which takes place at the $K'$ and $K$ points for $h^z=\frac{3J'}{7 g_z}$ and $h^z=-\frac{3J'}{7 g_z}$, respectively. (d) Above $h_c$ the quartet excitations become topologically trivial. For more information about the skyrmion of $\boldsymbol{\delta}_{\bf k}$ see SM~\ref{sec:skyrm}. ($J=1$, $J'=0.3$, $D_z=-0.3$, and $g_z=2$)}
\label{fig:topotrans_3o2}
\end{figure*}
%
The reason is that the original Hamiltonian~\ref{eq:Hamilton} contains only spin-spin interactions, and higher order, quadrupole-quadrupole or octupole-octupole terms are not present.

Naturally, both  $\boldsymbol{s}$ and $\boldsymbol{S}$ satisfy SU(2) algebra. Therefore, we can think about the multiplets as a spin-$S$ object in momentum space that is coupled to the pseudo magnetic field $\boldsymbol{d}^{(S)}_{\bf k}$. 
Diagonalizing Eqs.~\ref{eq:dirac} the eigenvalue of the $m$-th band in the $S$-multiplet has the form
$\omega_m({\bf k})=\Delta^{(S)}+m\cdot d^{(S)}_{\bf k}$, with $d^{(S)}_{\bf k}=|\boldsymbol{d}^{(S)}_{\bf k}|$.

It is apparent, that when $\boldsymbol{d}^{(S)}_{\bf k}$ vanishes at some point in the BZ, the bands with different $m$ touch at a single point. In case of the quartet, this corresponds to a novel spin-$3/2$ Dirac cone, with four bands touching at the transition point between topologically trivial and nontrivial phases. When $\boldsymbol{d}^{(S)}_{\bf k}$ is finite everywhere in the BZ, the multiplets are split and their topological properties can be characterized by a well-defined Chern number.
The pseudo magnetic field, $\boldsymbol{d}^{(S)}_{\bf k}$, carries all the topological information; When the bands are topologically non-trivial, $\boldsymbol{d}^{(S)}_{\bf k}$ forms a skyrmion in the BZ as illustrated above the dispersions in Figs.~\ref{fig:topo_1o2} and~\ref{fig:topotrans_3o2}. The skyrmion number $N_S$ is closely related to the Cern number of the bands which is $C^{(S)}_m=2mN_S$ with $m=-S,\hdots,S$ for the spin-$S$ multiplet~\cite{Romhanyi}.

We can use the magnetic field as tuning parameter to change the topological properties of the bands. 
At zero field, both $\boldsymbol{d}^{(1/2)}_{\bf k}$ and $\boldsymbol{d}^{(3/2)}_{\bf k}$ vanish at the $\gamma$ point, and the bands touch in both the spin-$1/2$ and spin-$3/2$ subspaces.
Switching on a tiny magnetic field the bands acquire topologically nontrivial characters without changing the chiral magnetic ground state as long as $h^z<h_{\rm pl}$.

We can formulate the topological properties using geometry. Drawing the $\boldsymbol{d}^{(S)}_{\bf k}$  vectors in a common origin, they map out a two-dimensional surface. When the origin, a monopole for Berry phase, is enclosed in the surface, the bands are topologically nontrivial, and when it is outside of the surface they are trivial. If $\boldsymbol{d}^{(S)}_{\bf k}=\boldsymbol{0}$  at a given point in the BZ, the origin is a point of the surface and the multiplet excitations are degenerate.  

The Dirac-like physics of the lower doublet excitations is shown in Fig.~\ref{fig:topo_1o2}. The dispersion of the spin-$1/2$ bands, $\omega^{(1/2)}_{\bf k}=\Delta^{(1/2)}\pm1/2\cdot d^{(1/2)}_{\bf k}$ , along with the surface formed by $\boldsymbol{d}^{(1/2)}_{\bf k}$ is shown for different values of magnetic field. Above the dispersion we plot the skyrmion formed by the pseudo-field $\boldsymbol{d}^{(1/2)}_{\bf k}$.
The $\boldsymbol{d}^{(1/2)}_{\bf k}$ and $\Delta^{(1/2)}_{\bf k}$ have the following form:

\begin{eqnarray}
&&d^{(1/2)}_x\!=\!\sum^3_{n=1}\frac{2J'}{9}\sqrt{1\!-\!\frac{(g_zh^z)^2}{J'^2}}\cos(\varphi+\alpha_n)(2-\cos(\boldsymbol{\delta}_n\boldsymbol{k}))\nonumber\\
&&d^{(1/2)}_y\!=\!\sum^3_{n=1}\frac{2J'}{9}\sqrt{1\!-\!\frac{(g_zh^z)^2}{J'^2}}\sin(\varphi+\alpha_n)(2-\cos(\boldsymbol{\delta}_n\boldsymbol{k}))\nonumber\\
&&d^{(1/2)}_z\!=\!\sum^3_{n=1}-\frac{2g_zh^z}{9}(\cos(\boldsymbol{\delta}_n\boldsymbol{k})\!+\!1)\!+\!\frac{\!\sqrt{3}\!}{9}J'\sin(\boldsymbol{\delta}_n\boldsymbol{k})
\label{eq:d_doublet}
\end{eqnarray}

\begin{eqnarray}
\Delta^{(1/2)}\!=\!\sum^3_{n=1}\frac{J'}{9}\!-\!\frac{\sqrt{3}}{3}D\!+\!\frac{\sqrt{3}}{18}g_zh^z\sin(\boldsymbol{\delta}_n\boldsymbol{k})
\label{eq:delta_doublet}
\end{eqnarray}

The translations $\boldsymbol{\delta}_n$, $(n=1,2,3)$ take the previously defined values, and the corresponding phases $\alpha_1=0$, $\alpha_2=-\frac{2\pi}{3}$ and $\alpha_3=\frac{2\pi}{3}$ reflect the Kitaev-type directional dependence of the effective inter-trimer interaction also discussed in Ref.~\cite{Repellin2017}. We recall that $\varphi=-5\pi/6$.

To see whether there is a topological transition in the doublet bands via closing and reopening the gap, we need to solve $\boldsymbol{d}^{(1/2)}_{\bf k}=0$. This is satisfied at the $K/K'$ points for field values $h^{(1/2)}_{\sf c}=\pm\frac{3J'}{2g_z}$. Therefore, at $h_c$, a band touching topological transition would occur at the $K/K'$ points. However, $h_c$ is larger than $h_{\sf plat}$, therefore the spin-$1/2$ multiplet remains topologically nontrivial all the way to the phase transition into the $1/3$-plateau phase.

The high-energy quartet splits too when the magnetic field  is finite and the four bands become topologically nontrivial. The surface traced out by $\boldsymbol{d}^{(3/2)}_{\bf k}$ is similar to what we found for the doublet, but here the orientation is reversed as illustrated with the reversed colors in Fig.~\ref{fig:topotrans_3o2}.  The extent of this surface is smaller. Consequently, the band touching topological transition occurs at lower magnetic fields, before the phases transition to the $1/3$-plateau phase. 
The explicit forms of $\boldsymbol{d}^{(3/2)}_{\bf k}$ and $\Delta^{(3/2)}_{\bf k}$ are
\begin{eqnarray}
&&d^{(3/2)}_x\!=\!\sum^3_{j=1}\textstyle{-\frac{J'}{9}\sqrt{1\!-\!\frac{(g_zh^z)^2}{J'^2}}\cos(\varphi-\alpha_n)(1+\cos(\boldsymbol{\delta}_n\boldsymbol{k}))}\nonumber\\
&&d^{(3/2)}_y\!=\!\sum^3_{j=1}\textstyle{-\frac{J'}{9}\sqrt{1\!-\!\frac{(g_zh^z)^2}{J'^2}}\sin(\varphi\!-\!\alpha_n)(1\!+\!\cos(\boldsymbol{\delta}_n\boldsymbol{k}))}\nonumber\\
&&d^{(3/2)}_z\!=\!\sum^3_{j=1}-\frac{g_zh^z}{18}(4\!+\!\cos(\boldsymbol{\delta}_n\boldsymbol{k}))\!-\!\frac{\!\sqrt{3}\!}{18}J'\sin(\boldsymbol{\delta}_n\boldsymbol{k})
\end{eqnarray}

\begin{eqnarray}
\Delta^{(3/2)}\!=\!\sum^3_{j=1}&&\frac{J}{2}\!-\!\frac{\sqrt{3}}{6}D\!+J'\!\frac{4\!-\!3\!\cos(\boldsymbol{\delta}_n\boldsymbol{k})\!}{36}\nonumber\\
&&-\frac{\sqrt{3}\!g_zh^z\!\sin(\boldsymbol{\delta}_n\boldsymbol{k})}{12}
\end{eqnarray}
with similar Kitaev-type directional dependent hopping than in the case of the lower doublet subspace.

Solving $\boldsymbol{d}^{(3/2)}_{\bf k}={\bf 0}$, we get a critical field $h^{(3/2)}_{\sf c}=\pm\frac{3J'}{7g_z}$ at which the topologically nontrivial bands form a novel spin-$3/2$ Dirac cone with four bands touching at the $K/K'$ point in the BZ as shown in Fig.~\ref{fig:topotrans_3o2} (c).  Above $h^{(3/2)}_{\sf c}$ the gap opens up again and the bands become topologically trivial. 
The vectors $\boldsymbol{d}^{(3/2)}_{\bf k}$ too form a skyrmion in the BZ with positive  or negative flux for $-\frac{3J'}{7 g_z}<h^z<0$ and $0<h^z<\frac{3J'}{7 g_z}$ field values, respectively.  
 The band-touching transition of the spin-$3/2$ bands is shown in Fig.~\ref{fig:topotrans_3o2}.
SM~\ref{sec:effective_1o2} and~\ref{sec:effective_3o2} contains details on the derivation of the pseudo-fields  $\boldsymbol{d}^{(S)}_{\bf k}$ and SM~\ref{sec:skyrm} on the skyrmions.

 \subsection{Multipolar edge states}


\begin{figure}[ht!]
\includegraphics[width=0.9\columnwidth]{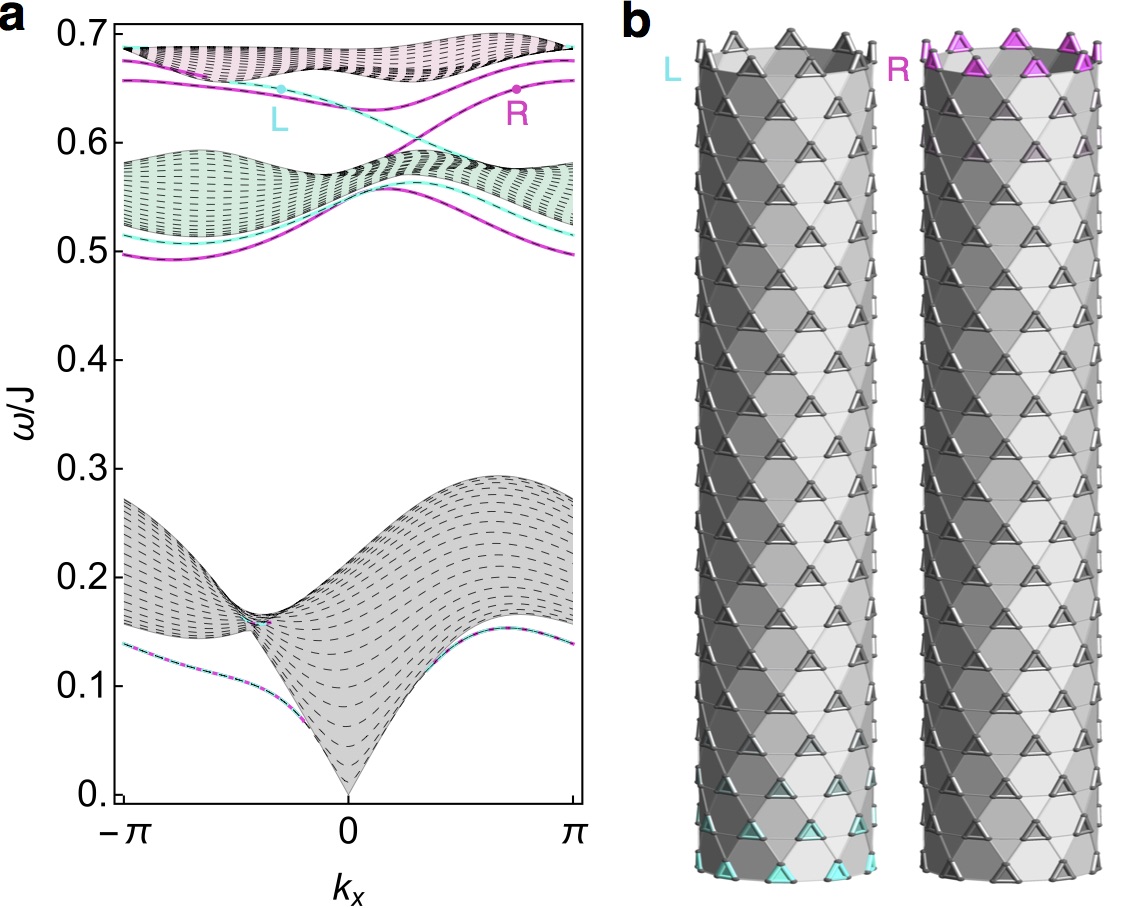}
\caption{Edge states appearing in an open system. The lattice has finite width in the $y$-direction and is periodic in the $x$-direction. (a) In the topological regime, in-gap edge states appear connecting the nontrivial bands of the low-energy part.  The projection of the bulk bands onto the $k_x$-axis is represented by colored regions.  The edge states connecting the Chernful bands are colored according to which side of the stripe they are localized. Panel (b) illustrates the exponential decay of the edge modes into the bulk with the same color-coding to distinguish the localization on opposite sides of the system at given energies as indicated in (a) with the points L and R. The magnetic field is $h^z=h_{\rm pl}/4 $ and the width of the stipe is 20 triangles.
}
\label{fig:edges_low}
\end{figure}


\begin{figure}[ht!]
\includegraphics[width=0.9\columnwidth]{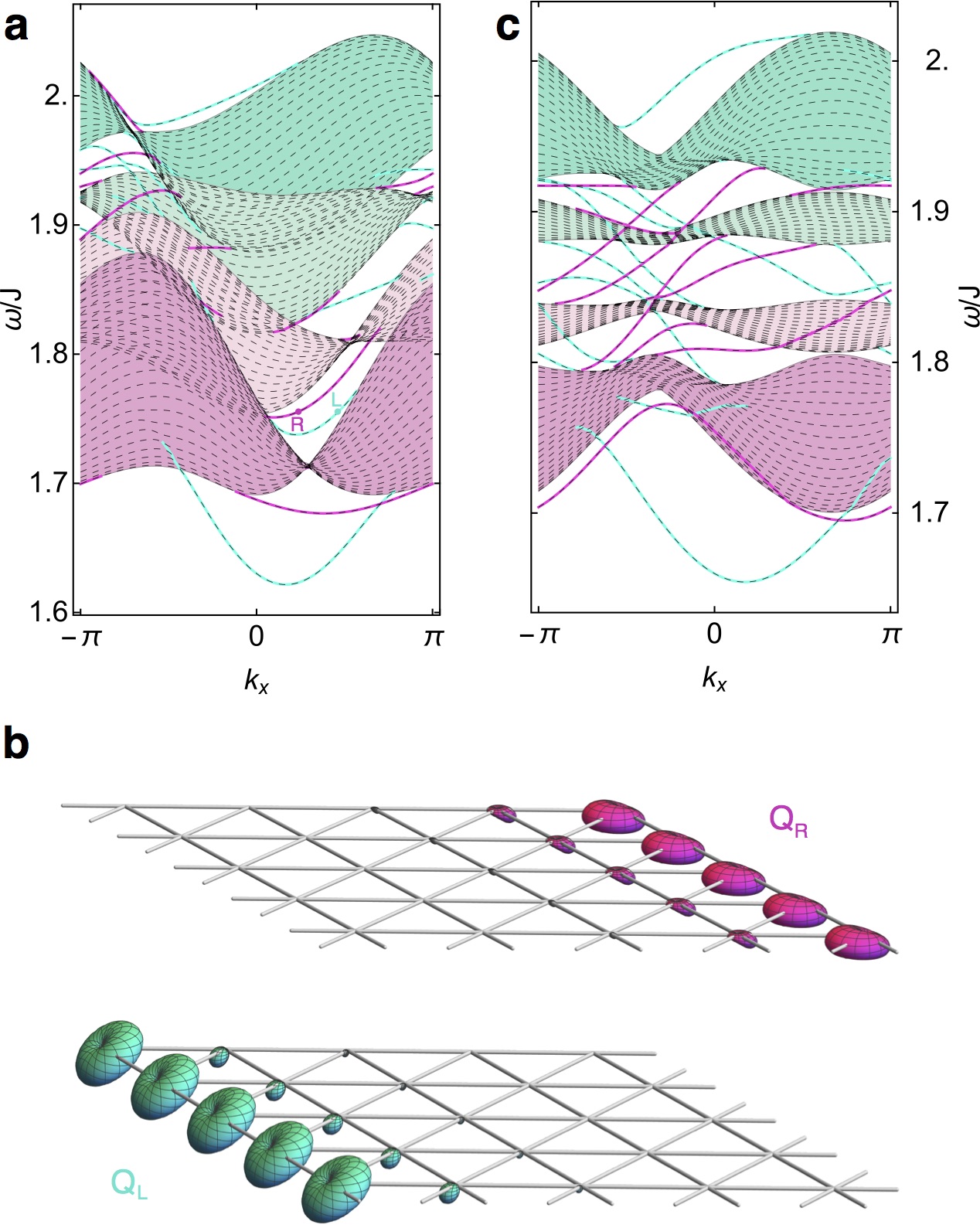}
\caption{Edge states in the high-energy quartet at $h^z=h_{\rm pl}/4 $. The lattice is open in the $y$- and periodic in the $x$-direction. (b) Spin polarized representation of the representative states R and L. The multipole degrees of freedom allowed by the larger Hilbert space of the $S=3/2$ quartet is naturally present in the edge states as well. For example, the two modes, R and L, correspond to predominantly quadrupolar excitations traveling on the right (Q$_{\sf R}$) and left (Q$_{\sf L}$) edges, respectively. The projection of the bulk bands is again represented by colored regions.  Due to the 'overlap' of the projected bands, it is difficult to keep track of the edge modes. To be able to see them we flatten the bands by continuously changing the model's parameters so that the topology is not affected as shown in panel (c). We count 3, 4 and 3 modes on each sides of the system in the three gaps separating the Chern bands. 
}
\label{fig:edges_high}
\end{figure}
%

The topological nature of bands can be shown through the appearance of edge states in an open system. As an example, we take periodic boundaries along the $x$-direction and open boundaries in the $y$-direction. The multiplet bands for this mixed boundary configuration are plotted in Fig.~\ref{fig:edges_low} and~\ref{fig:edges_high} together with the recovered bulk bands indicated with colored regions. 

In the stripe geometry, we find edge states in the bulk gap connecting the topologically nontrivial bands. To illustrate that these in-gap states are localized on the edges, we plot their exponential decay  into the bulk in Fig.~\ref{fig:edges_low}(b) for the doublet. In this case, there are two edge modes; one on each side of the sample, carrying only dipolar degrees of freedom, permitted for a $S=1/2$. 

The case of quartet multiplet is more interesting. The local Hilbert space of a spin $S=3/2$ allows for higher order multiplets, such as quadrupole and octupole degrees of freedom.  Thus, these multipole characters are naturally present in the quartet bands, promoting fundamentally new protected edge modes. 
In Fig.~\ref{fig:edges_high}(b) we plotted the representation of spin-component distribution for two given edge states indicated by the points L and R in Fig.~\ref{fig:edges_high}(a). 
The spin coherent state for a spin-$S$ can be written as
\begin{eqnarray}
\left|\boldsymbol{\Omega}\right>= \sum^S_{m=-S}e^{-i m\varphi}\sqrt{\binom{\!2S\!}{\!S\!+\!m\!}}\cos\!^{(\!S\!+\!m\!)}\!\frac{\vartheta}{2}\!\sin^{(\!S\!-\!m\!)}\!\frac{\vartheta}{2}\!\left|m\right>\nonumber\\
\end{eqnarray}
and the spin-component distribution of state $\left|\Psi\right>$ corresponds to the amplitude $\left|\left<\boldsymbol\Omega|\Psi\right>\right|^2$. In the stripe geometry the basis corresponds to $(\Psi_{1,k_x},\Psi_{2,k_x},\hdots,\Psi_{w,k_x})$, where $w$ is the finite width in $y$-direction. The $\Psi_{j,k_x}$ represents a state in the four dimensional local Hilbert space of the $j$th spin-$3/2$. In Fig.~\ref{fig:edges_high}(b) the amplitudes $\left|\left<\boldsymbol\Omega|\Psi_{j,k_x}\right>\right|^2$ are plotted for a selected $k_x$ point in the BZ to show not only the weight distribution of the edge states but also their spin-component distribution. 

In order to keep account of the edge states appearing in the quartet sub-space we smoothened the bands by changing parameters in the model in a way that the topology is not altered. When the bands are not covering each other, as shown in Fig.~\ref{fig:edges_high}(c), we can count the edge states. In the three gaps separating the four bands we find 3, 4 and 3 edge state on the left side with negative velocities and the same number of edge states on the right side with positive velocities, in agreement with the Chern numbers, -3,-1,1 and 3. 

\section{Discussion}

In summary, we studied a  relevant spin model of the trimerized kagom\'e lattice taking the entangled trimers as building units. The larger local Hilbert space characterizing the trimers naturally allows for multiplet excitations. As a result, an effective spin-$1/2$ and spin-$3/2$ band is formed above the gapless magnon mode. 

In the chiral magnetic state a small magnetic field removes the degeneracy of the multiplets and the excitations become topologically nontrivial with Chern numbers $C_m = 2m$. 
Increasing the magnetic field, the bands of the quartet undergo a topological transition when a novel spin-3/2 Dirac cone is formed by the touching of four bands. Above the critical field the bands become trivial. Before a similar topological transition occurred for the lower spin-1/2 doublet, at a higher magnetic field value, a conventional symmetry breaking transition takes place to the $1/3$-plateau phase, in which all bands are topologically trivial.

The spin-$1/2$ doublet carries only dipolar degrees of freedom, providing an analogue to the electronic systems. The  spin-$3/2$ quartet, on the other hand, encompass higher multipolar characters which in the topologically nontrivial regime is manifested in novel multipolar edge states.
Topologically robust edge states traveling unimpeded at the boundaries are appealing for low-energy consuming fast spintronical devices. Multipolar edge states, put forward here, may be the stepping stone for new directions in these endeavors.
A quadrupolar edge mode, for example, can couple to electric field and open a new route towards electric access and control of edge states emerging in the excitation spectrum of magnetic insulators. 

We expect that multipolar topological edge modes are present in a broad family of two-dimensional insulating quantum magnets, where the magnetic units are formed by larger spins, or entangled dimers or plaquettes. 
When the propagating spin-$S$ excitations acquire nontrivial topology, due to their multiplet properties, they exhibit novel spin-$S$ Dirac cones and large $-2S, \hdots, 2S$ Chern numbers.  Correspondingly, in open systems, chiral edge states emerge with inherent multipolar characters. 

\section{Acknowledgements}

The author is gratefully acknowledging discussions with Ludovic Jaubert, Karim Essafi, Nic Shannon and Karlo Penc. This work was supported by the Theory of Quantum Matter Unit of the Okinawa Institute of Science and Technology Graduate University (OIST) and the Hungarian OTKA Grant No. K 124176.

\section{Authors Contribution}
The author contributed to all aspects of this work.

\bibliographystyle{apsrev4-1}
\bibliography{Topo_Kagome}

\onecolumngrid

\section*{Supplementary Material:\\Multipolar edge states in the breathing kagom\'e model}


\renewcommand{\theequation}{S.\arabic{equation}}
\renewcommand{\thefigure}{S.\arabic{figure}}
\renewcommand{\thetable}{S.\Roman{table}}

\setcounter{section}{0}
\setcounter{equation}{0}
\setcounter{figure}{0}
\setcounter{table}{0}

\section{Ground state properties and multiplet basis}\label{sec:basis}

\begin{table}[h]\caption{The multiplet basis suitable for the model Hamiltonian~\ref{eq:Hamilton}. $S$ and $m$ characterize the SU(2) multiplet and its $z$ component, $\chi$ and $\chi^z_v$ are the scalar and vector chirality, respectively, and $\omega=\frac{2\pi}{3}$. This basis diagonalizes the intra-trimer part, and one of the four states, $\left|S=\frac 1 2,m,\chi^z_v\right>$ with $m=\pm\frac 1 2$ and $\chi^z_v=\pm $, is selected as a unique ground state depending on the signs of $h^z$ and $D$.}\label{tab:multiplets}
\begin{center}
\begin{tabular}{l@{\hspace{1cm}}cccc@{\hspace{1cm}}c}
\noalign{\vskip 0.1cm}
SU(2) & $S$ & $m$ & $\chi$ & $\chi^z_v$ & basis states  \\
\noalign{\vskip 0.1cm}
\hline\hline\noalign{\vskip 0.1cm}

\cmidrule(r){1-1}\cmidrule(lr){2-2}\cmidrule(l){3-3}
\multirow{2}{*}{$\mathcal{D}^{(1/2)}$} & \multirow{2}{*}{$\frac 1 2$} &$ \frac 1 2$ & + & + &$ -\frac{\duu+e^{i \omega}\udu+e^{-i\omega}\uud}{\sqrt{3}}$\\
&  &$ -\frac 1 2$ & + & -- & $ \frac{\udd+e^{i \omega}\dud+e^{-i\omega}\ddu}{\sqrt{3}}$\\
\noalign{\vskip 0.1cm}
\hline\noalign{\vskip 0.1cm}
\multirow{2}{*}{$\mathcal{D}^{(1/2)}$} & \multirow{2}{*}{$\frac 1 2$} &$ \frac 1 2$ & -- & -- &$ \frac{\duu+e^{-i \omega}\udu+e^{i\omega}\uud}{\sqrt{3}}$\\
&  &$ -\frac 1 2$ & -- & + &$ -\frac{\udd+e^{-i \omega}\dud+e^{i\omega}\ddu}{\sqrt{3}}$\\
\noalign{\vskip 0.1cm}
\hline\noalign{\vskip 0.1cm}
\multirow{4}{*}{$\mathcal{D}^{( 3/2)}$} & \multirow{4}{*}{$\frac 3 2$} &$ \frac 3 2$ & 0 & 0 & $\uuu$\\
& &$ \frac 1 2$ & 0 & 0 &$ \frac{\duu+\udu+\uud}{\sqrt{3}}$\\
&   &$- \frac 1 2$ & 0 & 0 & $ \frac{\udd+\dud+\ddu}{\sqrt{3}}$\\
&   &$ -\frac 3 2$ & 0 & 0 & $\ddd$\\
\noalign{\vskip 0.1cm}
\hline\hline
\end{tabular}
\end{center}
\end{table}
We collected the multiplet basis that diagonalizes the intra-trimer Hamiltonian in Table~\ref{tab:multiplets}. 
For finite magnetic field and DM interaction all degeneracy is lifted, and the ground state of the strong triangle is uniquely determined by the signs of the field $h^z$ and $D$.

We choose $D<0$ that favors the states with positive $z$-component of vector chirality. When the magnetic field is zero the lowest state of the trimer plaquette is two-fold degenerate;  $\left|S=\frac 1 2,\frac 1 2,+\right>$ and $\left|S=\frac 1 2,-\frac 1 2,+\right>$. Including the inter-trimer Heisenberg interaction, the ground state can be described as a general linear combination within this two-dimensional subspace:
\begin{eqnarray}
\textstyle{
\cos\vartheta \left|\frac 1 2,\frac 1 2,+\right> +e^{i \varphi} \sin\vartheta \left|\frac 1 2 ,\!-\!\frac 1 2,+\right>}\;,
\label{eq:gs}
\end{eqnarray}
The trimer-factorized mean-field energy is given in Eq.~\ref{eq:mf_en} in the main text. It only depends on  the parameter $\vartheta$ and is independent of the phase $\varphi$. This freedom in choosing $\varphi$ reflects the remaining U(1) symmetry of the system after the inclusion of DM interaction. 
Minimizing Eq.~\ref{eq:mf_en}, we get $\vartheta=\frac 1 2\arccos\frac{g_z h^z}{J'}$. The spins form a $120^\circ$ order that is canted out of the plane depending on the parameter $\vartheta$. 
 At zero field $\vartheta=\frac {\pi}{4}$,  the states mix equally and the spins lie in the lattice-plane, while at the critical field $h_c=J/g_z$, $\vartheta=0$ and a plateau phase is realized as the ground state with $m/m_{\rm sat}=1/3$. For convenience we choose $\varphi=-5\pi/6$. The spin expectation values then have the form of 
 \begin{eqnarray}
&& {\boldsymbol S}_1=\left(-\frac{1}{2\sqrt{3}}\sin2\vartheta,-\frac{1}{6}\sin2\vartheta,\frac{1}{6}\cos2\vartheta\right)\nonumber\\
&& {\boldsymbol S}_2=\left(\frac{1}{2\sqrt{3}}\sin2\vartheta,-\frac{1}{6}\sin2\vartheta,\frac{1}{6}\cos2\vartheta\right)\nonumber\\
&& {\boldsymbol S}_3=\left(0,\frac{1}{3}\sin2\vartheta,\frac{1}{6}\cos2\vartheta\right)
\label{eq:spins}
 \end{eqnarray}
 which correspond to the above described $120^\circ$ order which is canted out of the plane with $S^z_i=\frac{1}{6}\cos2\vartheta$. In the main text we use $\left|\allout\right>$ to denote this state.
 The state orthogonal to Eq.~\ref{eq:gs} plays the role of a pseudo-spin flip. It has the form  
 \begin{eqnarray}
\textstyle{\sin\vartheta \left|\frac 1 2,\frac1 2,+ \right> -e^{i \varphi} \cos\vartheta \left|\frac 1 2,\!-\!\frac 1 2,+\right>}\;,
\label{eq:exc}
\end{eqnarray}
and spin expectation values similar to Eq.~\ref{eq:spins} with opposite signs corresponding to an all-in $120^\circ$ configuration canted out of the plane with $S^z_i=-\frac{1}{6}\cos2\vartheta$. Therefore, in the main text we use the notation $\left|\allin\right>$ to refer to this state. 
The lowest excitation can be envisaged as the creation of this all-in state and its propagation on the lattice.
As discussed in the main text, we introduce a boson to create such pseudo-spin flip: $a^\dagger_j\left|0\right>$, where $\left|0\right>$ corresponds to the ground state, i.e. the condensation of $\left|\allout\right>$ which is the vacuum of the all-in triangles. 
We consider $\left|\allout\right>$ and $\left|\allin\right>$ as the two components of a pseudo spin-half, and 
project the original Hamiltonian~\ref{eq:Hamilton} onto this two-dimensional subspace. Using the well-known Holstein-Primakoff transformation, we write up the spin wave Hamiltonian of the $a^\dagger_j$ boson that creates the excitation $\left|\allin\right>$.
\begin{eqnarray}
\mathcal{H}^{\rm SW}&=&\sum_{\bf r} g_z h^z \cos2\vartheta a^{\dagger}_{\bf r}a^{\phantom{\dagger}}_{\bf r}+\sum_{\bf r} \sum_{n=1}^3\!\frac{J'}{36}(1-2\cos4\vartheta)(a^{\dagger}_{\bf r}a^{\phantom{\dagger}}_{\bf r}+a^{\dagger}_{{\bf r}+{\boldsymbol\delta}_n}a^{\phantom{\dagger}}_{{\bf r}+{\boldsymbol\delta}_n})\nonumber\\
&\!+\!&\!\sum_{\bf r} \sum_{n=1}^3\!\frac{J'}{72}(\!-\!5\!+\!i8\sqrt{3}\cos2\vartheta\!-\!3\cos4\vartheta)a^{\dagger}_{\bf r}a^{\phantom{\dagger}}_{{\bf r}+{\boldsymbol\delta}_n}+{\rm h.c.}\nonumber\\
&\!+\!&\sum_{\bf r} \sum_{n=1}^3\frac{J'}{12}\sin^2(2\vartheta)a^{\dagger}_{\bf r}a^{\dagger}_{{\bf r}+{\boldsymbol\delta}_n}+{\rm h.c.}
\end{eqnarray}
The position vector ${\bf r}$ runs over the unit cells and the translation vectors $\boldsymbol{\delta}_n$ can take the values listed in Tab.~\ref{tab:directions_dependence} below, and shown in Fig.~\ref{fig:lattice} in the main text. After Fourier transformation we retain the spin wave Hamiltonian Eq.~\ref{eq:lsw} of the main text which can be easily diagonalized using Bogoliubov transformation. 
 
 \section{Effective 2 by 2 hopping matrix for the spin-half multiplet}\label{sec:effective_1o2}

\begin{table}[h]\caption{Directional dependence of the neighboring bonds. Depending on the direction of the hopping process, the hopping amplitude acquires a phase factor $e^{i\alpha_n}$ introducing a Kitaev-like degrees of freedom. The bond $(i-j)$ denotes the site indices of the given bond $\boldsymbol{S}_{{\bf r},i}\cdot\boldsymbol{S}_{{\bf r}+\boldsymbol{\delta}_n,j}$.}\label{tab:directions_dependence}
\begin{center}
\begin{tabular}{l@{\hspace{1cm}}l@{\hspace{1cm}}l@{\hspace{1cm}}l}
\noalign{\vskip 0.1cm}
$n$ & bond & $\boldsymbol{\delta}_n$ & $\alpha_n$  \\
\noalign{\vskip 0.1cm}
\hline\hline\noalign{\vskip 0.1cm}
1 & (2-3) & $\boldsymbol{\delta}_1=\boldsymbol{\delta}_b-\boldsymbol{\delta}_a$ & $0$\\
2 & (3-1) & $\boldsymbol{\delta}_2=\boldsymbol{\delta}_a$ & $-\frac{2\pi}{3}$\\
3 & (1-2) & $\boldsymbol{\delta}_3=-\boldsymbol{\delta}_b$ & $\frac{2\pi}{3}$\\
\noalign{\vskip 0.1cm}
\hline\hline
\end{tabular}
\end{center}
\end{table}

Here we show the main steps for deriving the effective hopping Hamiltonian describing the dynamics of the low-lying doublet: $\left|\frac 1 2,+\frac 1 2,-\right>$ and $\left|\frac 1 2,-\frac 1 2,-\right>$.

We need to calculate matrix elements of the following form:
\begin{equation}
{}_i\langle \frac 1 2, m,-|\mathcal{H}|\frac 1 2,m', -\rangle_{j}
\end{equation}
where the site indices $i$ and $j$ are either the same or belong to neighboring up-triangles, and $m$ and $m'$ can take the values $-1/2$,  and $1/2$. 

We then introduce bosonic operators $\left|\frac 1 2,m,-\right>_{\bf r}=b^\dagger_{m,{\bf r}}\left|0\right>$  for the two states
and obtain the projected Hamiltonian to this 2 dimensional doublet sub-space:
\begin{eqnarray}
\mathcal{H}^{(1/2)}&=&\sum_{\bf r}\left[(-\sqrt{3}D-g_zh^z\sin^2\vartheta)b^\dagger_{\frac 1 2,{\bf r}}b^{\phantom\dagger}_{\frac 1 2,{\bf r}}+(-\sqrt{3}D+g_zh^z\cos^2\vartheta)b^\dagger_{\!-\!\frac 1 2,{\bf r}}b^{\phantom\dagger}_{\!-\!\frac 1 2,{\bf r}}\right]\nonumber\\
&\!+\!&\!\frac{1}{18}\!\sum_{\bf r}\sum_{n}\left[\sin^2\vartheta(2+3\cos2\vartheta)\left(b^\dagger_{\frac 1 2,{\bf r}}b^{\phantom\dagger}_{\frac 1 2,{\bf r}}+b^\dagger_{\frac 1 2,{\bf r}+\boldsymbol{\delta}_n}b^{\phantom\dagger}_{\frac 1 2,{\bf r}+\boldsymbol{\delta}_n}\right)\!+\!\cos^2\vartheta(2-3\cos2\vartheta)\left(b^\dagger_{\!-\!\frac 1 2,{\bf r}}b^{\phantom\dagger}_{\!-\!\frac 1 2,{\bf r}}+b^\dagger_{\!-\!\frac 1 2,{\bf r}+\boldsymbol{\delta}_n}b^{\phantom\dagger}_{\!-\!\frac 1 2,{\bf r}+\boldsymbol{\delta}_n}\right)\right]\nonumber\\
&+&\sum_{\bf r}\sum_{n}\frac{1}{36}\left[(i\sqrt{3}+(i\sqrt{3}-2)\cos2\vartheta)b^\dagger_{\frac 1 2,{\bf r}}b^{\phantom\dagger}_{\frac 1 2,{\bf r}+\boldsymbol{\delta}_n}+(-i\sqrt{3}+(i\sqrt{3}+2)\cos2\vartheta)b^\dagger_{\!-\!\frac 1 2,{\bf r}}b^{\phantom\dagger}_{\!-\!\frac 1 2,{\bf r}+\boldsymbol{\delta}_n}\right]+h.c.\nonumber\\
&+&\sum_{\bf r}\sum_{n}\frac{1}{9}e^{-i\varphi}e^{-i\alpha_n}\sin2\vartheta\left(b^\dagger_{\frac 1 2,{\bf r}}b^{\phantom\dagger}_{\!-\!\frac 1 2,{\bf r}}+b^\dagger_{\frac 1 2,{\bf r}+\boldsymbol{\delta}_n}b^{\phantom\dagger}_{\!-\!\frac 1 2,{\bf r}+\boldsymbol{\delta}_n}\right)+h.c.\nonumber\\
&-&\sum_{\bf r}\sum_{n}\frac{1}{18}e^{-i\varphi}e^{-i\alpha_n}\sin2\vartheta\left(b^\dagger_{\frac 1 2,{\bf r}}b^{\phantom\dagger}_{\!-\!\frac 1 2,{\bf r}+\boldsymbol{\delta}_n}+b^\dagger_{\frac 1 2,{\bf r}+\boldsymbol{\delta}_n}b^{\phantom\dagger}_{\!-\!\frac 1 2,{\bf r}}\right)+h.c.\;,
\label{eq:doublet_effective}
\end{eqnarray}
where ${\bf r}$ runs over the unit cells, and the sum for $n$ accounts for the three different directions of the three neighboring triangles. Due to the phase factor, $e^{-i\alpha_n}$, the hopping to the neighbors in the various directions is different, similarly to the Kitaev model. 
The bonds of different directions, and the corresponding translations, $\boldsymbol{\delta}_n$ and phase factors, $\alpha_n$ are collected in Tab.~\ref{tab:directions_dependence}.

After Fourier transformation, we can rewrite Eq.~\ref{eq:doublet_effective} as a two-by-two problem in the basis $(b^\dagger_{\frac{1}{2},{\bf k}}, b^\dagger_{\!-\!\frac 1 2,{\bf k}})$:

\begin{eqnarray}
\mathcal{H}^{(1/2)}=\sum_{\bf k}\sum_{n}
\left(
\begin{array}{c}
b^\dagger_{\frac{1}{2},{\bf k}}\\
b^\dagger_{\!-\!\frac{1}{2},{\bf k}}\\
\end{array}
\right)\left(
\begin{array}{cc}
A_{\bf k\phantom{\frac 1 2}} & C_{\bf k\phantom{\frac 1 2}}\\
C^*_{\bf k\phantom{\frac 1 2}} & B_{\bf k\phantom{\frac 1 2}}
\end{array}
\right)
\left(\begin{array}{c}
b^{\phantom\dagger}_{\frac{1}{2},{\bf k}}\\
b^{\phantom\dagger}_{\!-\!\frac{1}{2},{\bf k}}\\
\end{array}
\right)\label{eq:doublet_hop}
\end{eqnarray}
where 
\begin{eqnarray}
A_{\bf k}&=&\frac{J'}{9}-\frac{\sqrt{3}D}{3}-\frac{g_zh^z}{9}(1+\cos{\boldsymbol{\delta}}_n{\bf k})+\frac{\sqrt{3}}{18}\sin(\boldsymbol{\delta}_n{\bf k})(J'+g_zh^z)\;,\nonumber\\
B_{\bf k}&=&\frac{J'}{9}-\frac{\sqrt{3}D}{3}+\frac{g_zh^z}{9}(1+\cos{\boldsymbol{\delta}}_n{\bf k})+\frac{\sqrt{3}}{18}\sin(\boldsymbol{\delta}_n{\bf k})(-J'+g_zh^z)\;, \nonumber\\
C_{\bf k}&=&\frac{J'}{9}e^{-i\varphi}e^{-i\alpha_n}\textstyle{\sqrt{1-\left(\frac{g_zh^z}{J'}\right)^2}}(2-\cos(\boldsymbol{\delta}_n{\bf k}))
\end{eqnarray}
and we used the variational solution $\vartheta=\frac{1}{2}\arccos\frac{g_zh^z}{J'}$. The independent parameter $\varphi$ is chosen to be $-5\pi/6$.

We can bring Eq.~\ref{eq:doublet_hop} to the convenient form $\Delta^{(1/2)}_{\bf k}\cdot\boldsymbol{1}_{2}+\boldsymbol{d}^{(1/2)}_{\bf k}\cdot\boldsymbol{s}$,  as stated in the main text. The vector $\boldsymbol{s}$ is $\frac 1 2\boldsymbol{\sigma}$, with $\boldsymbol{\sigma}$ denoting the Pauli matrices. $\boldsymbol{1}_{2}$ is the 2-dimensional identity matrix and $\Delta^{(1/2)}_{\bf k}$ corresponds to the gap of the doublet.
The explicit forms of the vector $\boldsymbol{d}^{(1/2)}(\bf k)$ and $\Delta^{(1/2)}_{\bf k}$ are given in the main text in Eqs.~\ref{eq:d_doublet} and~\ref{eq:delta_doublet}, respectively. 

\section{Effective 4 by 4 hopping of the spin-$3/2$ quartet}\label{sec:effective_3o2}

We follow the same steps as in the case of the doublet in Sec.~\ref{sec:effective_1o2} and project the original problem onto the subspace of $\left|\frac 3 2,+\frac 3 2,0\right>$ $\left|\frac 3 2,+\frac 1 2,\right>$, $\left|\frac 3 2,-\frac 1 2,0\right>$ and $\left|\frac 3 2,-\frac 3 2,0\right>$. 
We need to calculate matrix elements of the following form:
\begin{equation}
{}_i\langle \frac 3 2, m,0|\mathcal{H}|\frac 3 2,m', 0\rangle_{j}
\end{equation}
where the indices $i$ and $j$ denote the sites of the same or neighboring up-triangles, furthermore $m$ and $m'$ can take the values $-3/2,\hdots,3/2$. 

Upon introducing bosonic operators $\left|\frac 3 2,m,0\right>_{\bf r}=c^\dagger_{m,{\bf r}}\left|0\right>$  for the four states we construct the projected hopping Hamiltonian of the quartet sub-space: $\mathcal{H}^{(3/2)}=\mathcal{H}^{(3/2)}_{\rm diag}+\mathcal{H}^{(3/2)}_{\rm off-diag}$

\begin{eqnarray}
\mathcal{H}^{(3/2)}_{\rm diag}&=&\sum_{\bf r}\textstyle{\left[\frac{3J}{2}-\frac{\sqrt{3}D}{2}+\frac{1}{2}g_zh^z\cos2\vartheta\right]}(c^\dagger_{\frac 3 2,{\bf r}}c^{\phantom\dagger}_{\frac 3 2,{\bf r}}+c^\dagger_{\frac 1 2,{\bf r}}c^{\phantom\dagger}_{\frac 1 2,{\bf r}}+c^\dagger_{\!-\!\frac 1 2,{\bf r}}c^{\phantom\dagger}_{\!-\!\frac 1 2,{\bf r}}+c^\dagger_{\!-\!\frac 3 2,{\bf r}}c^{\phantom\dagger}_{\!-\!\frac 3 2,{\bf r}})\nonumber\\
&+&\sum_{\bf r}g_zh^z\textstyle{\left[-\frac{3}{2} c^\dagger_{\frac 3 2,{\bf r}}c^{\phantom\dagger}_{\frac 3 2,{\bf r}}-\frac{1}{2}c^\dagger_{\frac 1 2,{\bf r}}c^{\phantom\dagger}_{\frac 1 2,{\bf r}}+\frac{1}{2}c^\dagger_{\!-\!\frac 1 2,{\bf r}}c^{\phantom\dagger}_{\!-\!\frac 1 2,{\bf r}}+\frac{3}{2}c^\dagger_{\!-\!\frac 3 2,{\bf r}}c^{\phantom\dagger}_{\!-\!\frac 3 2,{\bf r}}\right]}\nonumber\\
&\!+\!&\!\frac{1}{72}\!\sum_{\bf r}\sum_{n}(1+6\cos2\vartheta-3\cos4\vartheta)\left(c^\dagger_{\frac 3 2,{\bf r}}c^{\phantom\dagger}_{\frac 3 2,{\bf r}}+c^\dagger_{\frac 3 2,{\bf r}+\boldsymbol{\delta}_n}c^{\phantom\dagger}_{\frac 3 2,{\bf r}+\boldsymbol{\delta}_n}\right)\nonumber\\
&\!+\!&\!\frac{1}{18}\!\sum_{\bf r}\sum_{n}(2+3\cos2\vartheta)\sin^2\vartheta\left(c^\dagger_{\frac 1 2,{\bf r}}c^{\phantom\dagger}_{\frac 1 2,{\bf r}}+c^\dagger_{\frac 1 2,{\bf r}+\boldsymbol{\delta}_n}c^{\phantom\dagger}_{\frac 1 2,{\bf r}+\boldsymbol{\delta}_n}\right)\nonumber\\
&\!+\!&\!\frac{1}{18}\!\sum_{\bf r}\sum_{n}(2-3\cos2\vartheta)\cos^2\vartheta\left(c^\dagger_{\!-\!\frac 1 2,{\bf r}}c^{\phantom\dagger}_{\!-\!\frac 1 2,{\bf r}}+c^\dagger_{\!-\!\frac 1 2,{\bf r}+\boldsymbol{\delta}_n}c^{\phantom\dagger}_{\!-\!\frac 1 2,{\bf r}+\boldsymbol{\delta}_n}\right)\nonumber\\
&\!+\!&\!\frac{1}{72}\!\sum_{\bf r}\sum_{n}(1-6\cos2\vartheta-3\cos4\vartheta)\left(c^\dagger_{\!-\!\frac 3 2,{\bf r}}c^{\phantom\dagger}_{\!-\!\frac 3 2,{\bf r}}+c^\dagger_{\!-\!\frac 3 2,{\bf r}+\boldsymbol{\delta}_n}c^{\phantom\dagger}_{\!-\!\frac 3 2,{\bf r}+\boldsymbol{\delta}_n}\right)\nonumber\\
&\!+\!&\!\frac{1}{6}\!\sum_{\bf r}\sum_{n}\left[\cos^2\vartheta e^{-i\frac{2\pi}{3}}c^\dagger_{\frac 3 2,{\bf r}}c^{\phantom\dagger}_{\frac 3 2,{\bf r}+\boldsymbol{\delta}_n}+\sin^2\vartheta e^{i\frac{2\pi}{3}}c^\dagger_{\!-\!\frac 3 2,{\bf r}}c^{\phantom\dagger}_{\!-\!\frac 3 2,{\bf r}+\boldsymbol{\delta}_n}+h.c.\right]\nonumber\\
&\!+\!&\frac{1}{72}\sum_{\bf r}\sum_{n}\left[(-3-\cos2\vartheta)+i\sqrt{3}(-1-3\cos2\vartheta)\right]c^\dagger_{\frac 1 2,{\bf r}}c^{\phantom\dagger}_{\frac 1 2,{\bf r}}+h.c.\nonumber\\
&\!+\!&\frac{1}{72}\sum_{\bf r}\sum_{n}\left[(-3+\cos2\vartheta)+i\sqrt{3}(1-3\cos2\vartheta)\right]c^\dagger_{\!-\!\frac 1 2,{\bf r}}c^{\phantom\dagger}_{\!-\!\frac 1 2,{\bf r}+\boldsymbol{\delta}_n}+h.c.
\label{eq:quartet_diag}
\end{eqnarray}

\begin{eqnarray}
\mathcal{H}^{(3/2)}_{\rm off-diag}&=&\frac{\sqrt{3}}{18}\sum_{\bf r}\sum_{n}e^{-i\varphi}e^{i\alpha_n}\sin2\vartheta\left[e^{i\frac{2\pi}{3}}c^\dagger_{\frac 3 2,{\bf r}}c^{\phantom\dagger}_{\frac 1 2,{\bf r}}+e^{-i\frac{2\pi}{3}}c^\dagger_{\frac 3 2,{\bf r}+\boldsymbol{\delta}_n}c^{\phantom\dagger}_{\frac 1 2,{\bf r}+\boldsymbol{\delta}_n}\right]+h.c.\nonumber\\
&+&\frac{1}{9}\sum_{\bf r}\sum_{n}e^{-i\varphi}e^{i\alpha_n}\sin2\vartheta\left[e^{i\frac{2\pi}{3}}c^\dagger_{\frac 1 2,{\bf r}}c^{\phantom\dagger}_{-\frac 1 2,{\bf r}}+e^{-i\frac{2\pi}{3}}c^\dagger_{\frac 1 2,{\bf r}+\boldsymbol{\delta}_n}c^{\phantom\dagger}_{-\frac 1 2,{\bf r}+\boldsymbol{\delta}_n}\right]+h.c.\nonumber\\
&+&\frac{\sqrt{3}}{18}\sum_{\bf r}\sum_{n}e^{-i\varphi}e^{i\alpha_n}\sin2\vartheta\left[e^{i\frac{2\pi}{3}}c^\dagger_{-\frac 1 2,{\bf r}}c^{\phantom\dagger}_{-\frac 3 2,{\bf r}}+e^{-i\frac{2\pi}{3}}c^\dagger_{-\frac 1 2,{\bf r}+\boldsymbol{\delta}_n}c^{\phantom\dagger}_{-\frac 3 2,{\bf r}+\boldsymbol{\delta}_n}\right]+h.c.\nonumber\\
&-&\frac{\sqrt{3}}{36}\sum_{\bf r}\sum_{n}e^{-i\varphi}e^{i\alpha_n}\sin2\vartheta\left[c^\dagger_{\frac 3 2,{\bf r}}c^{\phantom\dagger}_{\frac 1 2,{\bf r}+\boldsymbol{\delta}_n}+c^\dagger_{\frac 3 2,{\bf r}+\boldsymbol{\delta}_n}c^{\phantom\dagger}_{\frac 1 2,{\bf r}}\right]+h.c.\nonumber\\
&-&\frac{1}{18}\sum_{\bf r}\sum_{n}e^{-i\varphi}e^{i\alpha_n}\sin2\vartheta\left[c^\dagger_{\frac 1 2,{\bf r}}c^{\phantom\dagger}_{-\frac 1 2,{\bf r}+\boldsymbol{\delta}_n}+c^\dagger_{\frac 1 2,{\bf r}+\boldsymbol{\delta}_n}c^{\phantom\dagger}_{-\frac 1 2,{\bf r}}\right]+h.c.\nonumber\\
&-&\frac{\sqrt{3}}{36}\sum_{\bf r}\sum_{n}e^{-i\varphi}e^{i\alpha_n}\sin2\vartheta\left[c^\dagger_{-\frac 1 2,{\bf r}}c^{\phantom\dagger}_{-\frac 3 2,{\bf r}+\boldsymbol{\delta}_n}+c^\dagger_{-\frac 1 2,{\bf r}+\boldsymbol{\delta}_n}c^{\phantom\dagger}_{-\frac 3 2,{\bf r}}\right]+h.c.
\label{eq:quartet_off_diag}
\end{eqnarray}

The translations ${\boldsymbol\delta}_n$ and the phase factors $e^{-i\alpha_n}$ take the values defined in Tab.~\ref{tab:directions_dependence}. Similar to the doublet effective model, the hopping to the neighbors has Kitaev-like direction dependence through the phase factor $e^{-i\alpha_n}$. 

After Fourier transformation, we can rewrite Eqs.~\ref{eq:quartet_off_diag} and~\ref{eq:quartet_diag} as a four-by-four hopping Hamiltonian in the basis $(c^\dagger_{\frac{3}{2},{\bf k}}, c^\dagger_{\frac{1}{2},{\bf k}}, c^\dagger_{\!-\!\frac 1 2,{\bf k}}, c^\dagger_{\!-\!\frac3 2,{\bf k}})$:

\begin{eqnarray}
\mathcal{H}^{(1/2)}=\sum_{\bf k}\sum^3_{n=1}
\left(
\begin{array}{c}
c^\dagger_{\frac{3}{2},{\bf k}}\\
c^\dagger_{\frac{1}{2},{\bf k}}\\
c^\dagger_{\!-\!\frac{1}{2},{\bf k}}\\
c^\dagger_{\!-\!\frac{3}{2},{\bf k}}\\
\end{array}
\right)\left(
\begin{array}{cccc}
M_{\frac 3 2,\frac 3 2} & M_{\frac 3 2,\frac 1 2} & 0 & 0\\
M^*_{\frac 3 2,\frac 1 2} & M_{\frac 1 2,\frac 1 2} & M_{\frac 1 2,-\frac 1 2} & 0\\
0 & M^*_{\frac 1 2,-\frac 1 2} & M_{-\frac 1 2,-\frac 1 2} & M_{-\frac 1 2,-\frac 3 2}\\
0 & 0 & M^*_{-\frac 1 2,-\frac 3 2} & M_{-\frac 3 2,-\frac 3 2}
\end{array}
\right)
\left(\begin{array}{c}
c^{\phantom\dagger}_{\frac{3}{2},{\bf k}}\\
c^{\phantom\dagger}_{\frac{1}{2},{\bf k}}\\
c^{\phantom\dagger}_{\!-\!\frac{1}{2},{\bf k}}\\
c^{\phantom\dagger}_{\!-\!\frac{3}{2},{\bf k}}\\
\end{array}
\right)\label{eq:quartet_hop}
\end{eqnarray}
where the diagonal terms are
\begin{eqnarray}
&&M_{\frac 3 2,\frac 3 2}=\frac{J}{2}-\frac{D}{2\sqrt{3}}+\frac{J'}{9}-\frac{J'}{12}\cos(\boldsymbol{\delta}_n{\bf k})-\frac{J'}{4\sqrt{3}}\sin(\boldsymbol{\delta}_n{\bf k})-\frac{g_zh^z}{4\sqrt{3}}\sin(\boldsymbol{\delta}_n{\bf k})-\frac{g_zh^z}{12}(4+\cos(\boldsymbol{\delta}_n{\bf k}))\nonumber\\
&&M_{\frac 1 2,\frac 1 2}=\frac{J}{2}-\frac{D}{2\sqrt{3}}+\frac{J'}{9}-\frac{J'}{12}\cos(\boldsymbol{\delta}_n{\bf k})-\frac{J'}{12\sqrt{3}}\sin(\boldsymbol{\delta}_n{\bf k})-\frac{g_zh^z}{4\sqrt{3}}\sin(\boldsymbol{\delta}_n{\bf k})-\frac{g_zh^z}{36}(4+\cos(\boldsymbol{\delta}_n{\bf k}))\nonumber\\
&&M_{-\frac 1 2,-\frac 1 2}=\frac{J}{2}-\frac{D}{2\sqrt{3}}+\frac{J'}{9}-\frac{J'}{12}\cos(\boldsymbol{\delta}_n{\bf k})+\frac{J'}{12\sqrt{3}}\sin(\boldsymbol{\delta}_n{\bf k})-\frac{g_zh^z}{4\sqrt{3}}\sin(\boldsymbol{\delta}_n{\bf k})+\frac{g_zh^z}{36}(4+\cos(\boldsymbol{\delta}_n{\bf k}))\nonumber\\
&&M_{-\frac 3 2,-\frac 3 2}=\frac{J}{2}-\frac{D}{2\sqrt{3}}+\frac{J'}{9}-\frac{J'}{12}\cos(\boldsymbol{\delta}_n{\bf k})+\frac{J'}{4\sqrt{3}}\sin(\boldsymbol{\delta}_n{\bf k})-\frac{g_zh^z}{4\sqrt{3}}\sin(\boldsymbol{\delta}_n{\bf k})+\frac{g_zh^z}{12}(4+\cos(\boldsymbol{\delta}_n{\bf k}))
\end{eqnarray}
and the off-diagonal part is given by:
\begin{eqnarray}
&&M_{\frac 3 2,\frac 1 2}=-\frac{J\sqrt{3}}{18}e^{-i\varphi}e^{i\alpha_n}\sin2\vartheta(1+\cos(\boldsymbol{\delta}_n{\bf k}))\nonumber\\
&&M_{\frac 1 2,\frac 1 2}=-\frac{J}{9}e^{-i\varphi}e^{i\alpha_n}\sin2\vartheta(1+\cos(\boldsymbol{\delta}_n{\bf k}))\nonumber\\
&&M_{-\frac 1 2,-\frac 1 2}=-\frac{J\sqrt{3}}{18}e^{-i\varphi}e^{i\alpha_n}\sin2\vartheta(1+\cos(\boldsymbol{\delta}_n{\bf k}))
\end{eqnarray}
The free parameter $\varphi$ is again $-5\pi/6$ and we need to insert the variational solution, $\frac{1}{2}\arccos\frac{g_zh^z}{J'}$ for $\vartheta$.

We started with a purely spin-spin problem, and no higher order tensors, such as quadrupolar-quadrupolar terms were taken into account in the original Hamiltonian Eq.~\ref{eq:Hamilton} defined in the main text. For this reason the effective hopping model contains only spin-like terms, such as spin raising and lowering components and diagonal contributions, but there is no quadrupole or octupole components that were otherwise allowed by the larger Hilbert space of the effective spin-$3/2$ object. Consequently, the effective hopping model for the quartet~\ref{eq:quartet_hop} can also be written in the form of $\Delta^{(3/2)}_{\bf k}\cdot\boldsymbol{1}_{4}+\boldsymbol{d}^{(3/2)}_{\bf k}\cdot\boldsymbol{S}$, where the $\boldsymbol{1}_{4}$ is a 4-dimensional identity matrix, and $\boldsymbol{S}$ is the vector constructed from the spin operators of an $S=3/2$ quantum spin:

\begin{gather}
S^x\! =\!\!
\!\!\left(\begin{array}{cccc}
0 & \!\frac{\sqrt{3}}{2}\! & 0 & 0\\
\!\frac{\sqrt{3}}{2}\! & 0 & 1 & 0\\
0 & 1 & 0 &\!\frac{\sqrt{3}}{2}\!\\
0 & 0 & \!\frac{\sqrt{3}}{2}\! & 0
\end{array}\right)\!\!,\quad
S^y\! =\!\!
\!\!\left(\begin{array}{cccc}
0 & \!\frac{-i\sqrt{3}}{2}\! & 0 & 0\\
\!\frac{i\sqrt{3}}{2}\! & 0 & -i & 0\\
0 & i & 0 & \!\frac{-i\sqrt{3}}{2}\!\\
0 & 0 & \!\frac{i\sqrt{3}}{2}\! & 0
\end{array}\right)\!\!,\quad
S^z=
\!\!\left(\begin{array}{cccc}
\!\frac 3 2\! & 0 & 0 & 0\\
0 & \!\frac 1 2\! & 0 & 0\\
0 & 0 & \!-\frac {1}{ 2}\! & 0\\
0 & 0 & 0 & \!-\frac{3}{2}\!
\end{array}\right)\!\!\!
\end{gather}

\section{The forms of vector $\boldsymbol{d}^{(S)}_{\bf k}$ and the transition points}\label{sec:skyrm}

\begin{figure}[h!]
\includegraphics[width=0.8\columnwidth]{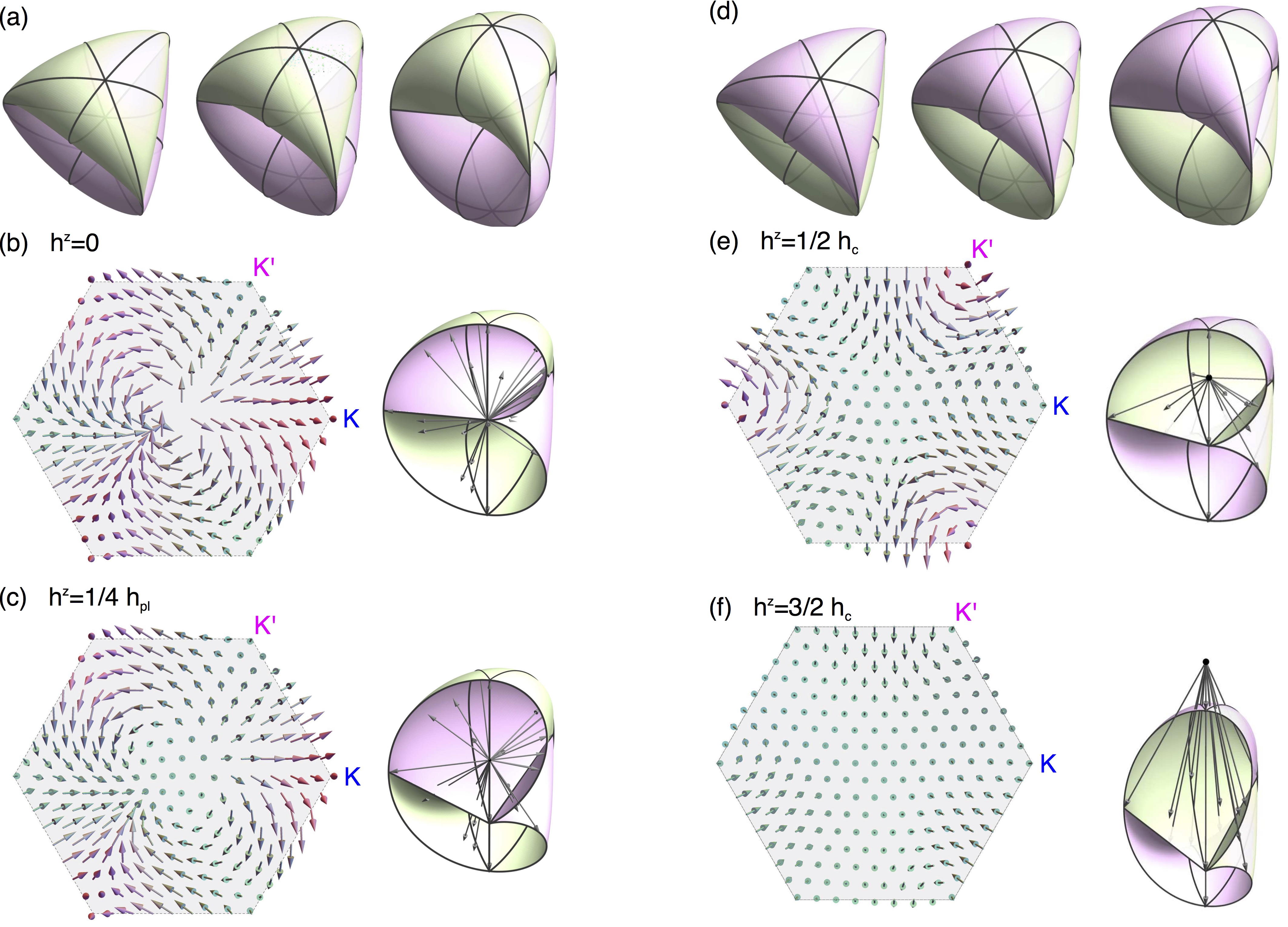}
\caption{
(a) and (d) show the two-dimensional surface spanned by the vectors $\boldsymbol{d}^{(1/2)}_{\bf k}$ and $\boldsymbol{d}^{(3/2)}_{\bf k}$, respectively. The different viewpoints help envisaging the 3-fold symmetry carried over from the lattice to the vectors $\boldsymbol{d}^{(S)}_{\bf k}$ and the the two chambers enclosed. When the origin of the vectors is inside one of the chambers, the corresponding multiplet bands are topologically non-trivial and the $\boldsymbol{d}^{(S)}_{\bf k}$ forms a skyrmion in the Brillouin zone as shown in (c) and (e). If the origin is a part of the surface, necessarily one of the $\boldsymbol{d}^{(S)}_{\bf k}$ vectors has zero length, and the multiplets are degenerate with undefined Chern numbers, such as the case for zero magnetic field as shown in (b) in case of the doublet. When the origin is outside of the surface, the bands are fully gapped again, with trivial topology and zero Chern numbers as illustrated in (f) for the quartet.
}
\label{fig:skyrmion}
\end{figure}

As discussed in the main tex the topology of the system is encompassed in the  pseudo magnetic fields $\boldsymbol{d}^{(S)}_{\bf k}$. In Fig.~\ref{fig:skyrmion} we illustrate this property. Subfigures~\ref{fig:skyrmion}(a), (b) and (c) belong to the doublet, while (c), (d) and (e) are representatives of the quartet. 
We plot the surface spanned by $\boldsymbol{d}^{(1/2)}_{\bf k}$ and $\boldsymbol{d}^{(3/2)}_{\bf k}$ in Fig.~\ref{fig:skyrmion}(a) and (d), respectively, from different point of views to help visualize the shape of the surface.
It has a 3-fold symmetry along the $z$-axis, inherited from the symmetry of the lattice. Furthermore, it encloses two separate chambers. Wen the origin of the $\boldsymbol{d}^{(S)}_{\bf k}$ vectors is inside, the bands are fully gapped and the spin-$S$ multiplet is topologically nontrivial with Chern numbers $-2S,\hdots, 2S$. In such cases the vectors $\boldsymbol{d}^{(S)}_{\bf k}$ form a skyrmion in the Brillouin zone (BZ) as shown in Fig.~\ref{fig:skyrmion}(c) and (e) for the doublet and quartet.
When the origin sits on the surface, a $\boldsymbol{d}^{(S)}_{\bf k}$ is zero somewhere in the BZ and the bands touch, forming a novel spin-$S$ Dirac cone. Such is the case for example at zero magnetic field shown in Fig.~\ref{fig:skyrmion}(b) for the doublet. In cases when the origin is outside the chambers, the bands are fully gapped and topologically trivial as indicated in Fig.~\ref{fig:skyrmion}(f). This only happens in the case of the quartet. The transition field for the doublet is too high for such a band touching topological transition, and at lower critical field a symmetry breaking phase transition takes place into the $1/3$-magnetization plateau phase.

\end{document}